\documentclass[a4paper,11pt]{article}
\pdfoutput=1 

\usepackage{jcappub} 

\usepackage[T1]{fontenc} 
\usepackage{xcolor}

\title{\boldmath Bayesian analysis for a class of $\alpha$-attractor inflationary models}

\author[a,1]{Francisco X. Linares Cede\~no,\note{Corresponding author.}}
\author[b]{Gabriel German,}
\author[b]{Juan Carlos Hidalgo,}
\author[c]{and Ariadna Montiel}

\affiliation[a]{Departamento Ingenier\'ia Civil, Divisi\'on de Ingenier\'ia, Universidad de Guanajuato, Guanajuato CP 36000, M\'exico}
\affiliation[b]{Instituto de Ciencias F\'{\i}sicas, Universidad Nacional Aut\'onoma de M\'exico, 62210, Cuernavaca, Morelos, M\'exico}
\affiliation[c]{Depto. de F\'isica y Matem\'aticas, Tecnol\'ogico de Monterrey, Campus Estado de M\'exico,
Atizap\'an de Zaragoza, Estado de M\'exico, Apdo. 52926, M\'exico}

\emailAdd{fran2012@fisica.ugto.mx}
\emailAdd{gageve@gmail.com}
\emailAdd{hidalgo@icf.unam.mx}
\emailAdd{amontiel@icf.unam.mx}

\abstract{We perform a Bayesian study of a generalization of the basic $\alpha$-attractor T model given by the potential $V(\phi)=V_0\left[1-\text{sech}^{p}\left(\phi/\sqrt{6\alpha}M_{pl}\right)\right]$ where $\phi$ is the inflaton field and the parameter $\alpha$ corresponds to the inverse curvature of the scalar manifold in the conformal or superconformal realizations of the attractor models. Such generalization is characterized by the power $p$ which includes the basic or base model for $p=2$. Once the priors for the parameters of the $\alpha$-attractor potential are set by numerical exploration, we perform the corresponding statistical analysis for the cases $p=1\, , 2\, , 3\, ,4$, and derive posteriors. Considering the original $\alpha$-attractor potential as the base model, we calculate the evidence for our generalization, and conclude that the $p=4$ model is preferred by the CMB data. We also present constraints for the parameter $\alpha$. Interestingly, all the cases studied prefer a specific value for the tensor-to-scalar ratio given by $r\simeq 0.0025$. }

\begin{document}
\maketitle
\flushbottom

\section{Introduction}
\label{sec:intro}

The physical processes that gave origin to our universe remain unknown. The most accepted hypothesis, that at the very beginning the spacetime undergoes an exponential expansion, is the so called  \textit{inflation paradigm}~\cite{Guth:1980zm,Linde:1981mu}. Once such expansion ends, a reheating period must have taken place leading to the hot big bang. Inflation is usually modelled by a scalar field, dubbed the \textit{inflaton} field. It is approximately after 380,000 years from this initial stage, that we are able to look at the early universe through the first light coming from the Cosmic Microwave Background radiation (CMB) and infer the physics of the inflationary era. Particularly, the primordial fluctuations of the inflaton field are imprinted in several observables, such as the spectral index $n_s$, the scalar and tensor amplitude $A_s$ and $A_t$ respectively, and the quotient between these quantities: the tensor-to-scalar ratio,
\begin{equation}
r = \frac{A_t}{A_s} .
\label{r} 
\end{equation}

This synthesizes in a single number the tensor fluctuations of the  metric field, also known as the \textit{primordial gravitational waves}. This is a prediction of the inflationary paradigm, and several experiments have been performed to detect them, such as Background Imaging of Cosmic Extragalactic Polarization (BICEP, $r<0.73$ at $95\%$)~\cite{Chiang:2009xsa}, QUEST at DASI (QUaD, $r<0.48$ at $95\%$)~\cite{QUaD:2009gka}, the Wilkinson Microwave Anisotropy Probe (WMAP, $r<0.38$ at $95\%$)~\cite{WMAP:2012fli}, the Q/U Imaging ExperimenT (QUIET, $r<2.7$ at $95\%$)~\cite{QUIET:2012szu}, Planck Satellite Collaboration ($r<0.06$ at $95\%$)~\cite{Planck:2018vyg}, the South Pole Telescope (SPTpol, $r<0.44$ at $95\%$)~\cite{SPT:2019nip}. The current upper bounds on $r$ have been reported by the combined BICEP/Keck and Planck data~\cite{Tristram:2021tvh}, with a value given by $r<0.032$ at $95\%$, whereas futures experiments such as LiteBIRD would impose constraints on the tensor-to-scalar ratio with upper limits  of the order $r \simeq 0.003$~\cite{LiteBIRD:2022cnt}. It is therefore of great relevance to investigate the dynamics of the inflaton field, and looking for some prediction of particular models with the aim of elucidate which kind of potentials are able to reproduce CMB observations.

From all the families of scalar field potentials (see e.g.~Table 1 from~\cite{Martin:2013tda} for some examples of several forms of $V(\phi)$), we will focus in those known as $\alpha$\textit{-attractor models}. Specifically, we carry out a Bayesian study of  $\alpha$-attractor models of inflation of the type T~\cite{Kallosh:2013yoa}. Several realizations of $\alpha$-attractors have been studied with considerable detail in recent years\footnote{Even the authors have studied such models in the context of dark energy~\cite{LinaresCedeno:2019bgo}.}~\cite{Akrami:2020zxw,Canas-Herrera:2021sjs,Sarkar:2021ird,Iacconi:2021ltm,Kallosh:2022feu,Krajewski:2022ezo,Braglia:2022phb}, since they fall at the so called \textit{sweet spot} of the parameter space preferred by {\sc Planck} data~\cite{Planck:2018jri}. However, to the best of our knowledge, a systematic Bayesian study of them is still missing. The present article intends to address this problem using a generalization~\cite{German:2021rin} $V =V_0\left[1-\text{sech}^{p}(\phi/\sqrt{6\alpha}M_{pl})\right]$  of the basic model $V=V_0\tanh^2(\phi/\sqrt{6\alpha}M_{pl})$ of $\alpha$-attractors of type T characterized by the presence of a definite positive function that allows the generalization to be viable as an inflation model for any rational value of the parameter $p$. This generalization also has the peculiarity that the first term in the expansion of the potential around the origin, where reheating should take place, is quadratic with the parameter $p$ appearing only as a multiplicative factor of $\phi^2$ and of higher order terms, where $\phi$ is the inflaton field. We focus in particular the cases with $p=1\, , 2\, , 3\, $ and $4$, which have been also studied analytically within the slow roll approximation~\cite{German:2020cbw}.

\begin{figure}[h!]
\begin{center}
\includegraphics[width=8cm, height=7.0cm]{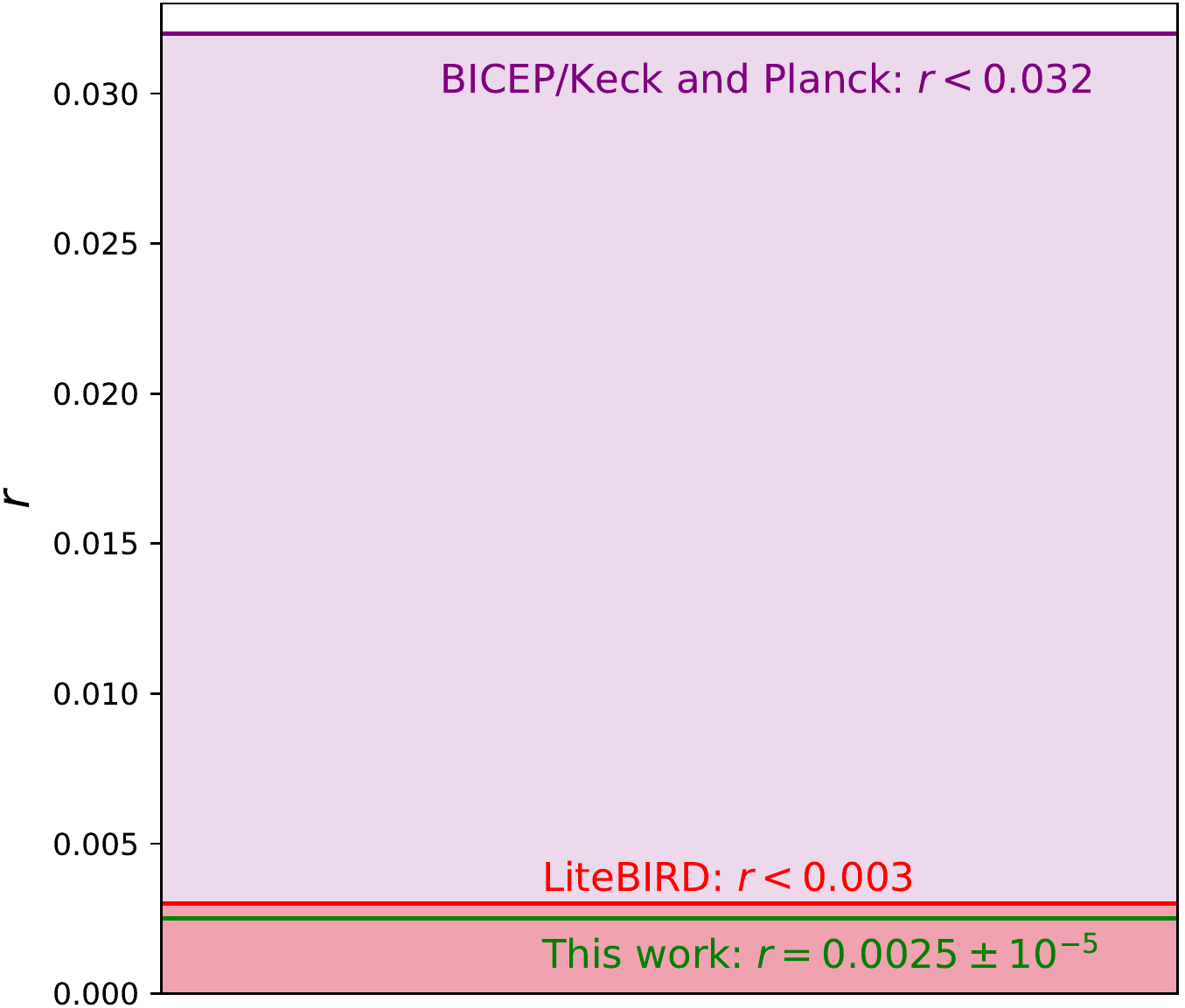}
\caption{\small Current and forecasted upper bounds imposed by BICEP/Keck and Planck (purple)~\cite{Tristram:2021tvh} and LiteBIRD (red)~\cite{LiteBIRD:2022cnt} collaborations for the tensor-to-scalar ratio $r$ (at 95\% confidence level). Our result (green) is in good agreement with such constraints. See text for more details.}
\label{r_our_results}
\end{center}
\end{figure}

In our analysis, the members of the family of potentials (models) are identified by a specific value of the exponent $p$ defined above, whereas the main parameters of the model to be constrained are the amplitude of the potential $V_0$, and $\lambda$ (in the original notation $\lambda \equiv1/\sqrt{6\alpha}$). This is directly related to the curvature of the inflaton scalar manifold. Specifically, the K$\ddot{{\rm{a}}}$hler metric defines the metric in the field space, and $\alpha$ (or $\lambda$) is inversely (directly) related to the curvature scalar in the inner space (for more details see~\cite{Kallosh:2013yoa,German:2019aoj}). Within the Bayesian study itself, we consider flat priors for $(V_0\, ,\lambda)$, which were settle by the aim of numerical exploration with the Boltzmann code \textsc{class}~\cite{Lesgourgues:2011re}, which incorporates a module to simulate inflation for a given inflationary scalar field potential $V(\phi)\, .$ The statistical analysis was carried out with the public Monte Carlo code \textsc{Monte Python}~\cite{Audren:2012wb}, from which the posteriors for the inflaton potential parameters, (the logarithm of) $V_0$ and $\lambda\, , $ are derived, as well as those of cosmological parameters such as the number of e-folds for the inflationary period $N_{\star}$, the spectral index $n_s$, and the tensor-to-scalar ratio $r$. The value of the latter is an important result: for any of the values of $p$ we have explored, all the cases present  $r\simeq 0.0025$. Figure~\ref{r_our_results} shows that this result is in good agreement with current and forecast constraints on $r$. As a final part of our analysis, we perform a Bayesian selection of models within the family under study with $p=1\, , 2\, , 3\, , 4$, for which we have used the public software \textsc{MCEvidence}~\cite{Heavens:2017afc}. Our results show that the case $p=4$ is favoured over the base model with $p=2$, which represents the standard $\alpha$-attractor model with tangent potential. On the other hand, the cases $p=1$ and $p=3$ are disfavoured by CMB data.

This article is organised as follows: in Section \ref{generalizations} we briefly review the basic  $\alpha$-attractor model of inflation of the type T with emphasis on two of its generalizations: One given by the potential of Eq.~\eqref{potanh} and a second generalization, the one here studied, given by the potential in \eqref{potsech}. Relevant formulas for the slow-roll parameters and observables are presented and relations between the parameters of the inflationary model with their values in the \textsc{class} package are established. In Section \ref{numerical} we perform a numerical analysis of the model, by including the potential \eqref{potsech} in the Boltzmann code \textsc{class}. This allows us to establish the priors for the potential parameters $V_0$ and $\lambda$, such that the range of values for both of them, for any value of $p=1\, , 2\, , 3\, , 4$, are consistent with reported values of observed physical quantities such as the scalar amplitude $A_s$, the spectral index $n_s$, and the tensor-to-scalar ratio $r$. In Section \ref{statistical} we show the results from the statistical analysis. We use the Bayes's Theorem: 1.- to obtain the posteriors for $[\log V_0\, , \lambda\, , N_{\star}\, , n_s\, , r]$ and 2.- to calculate the evidence and the Bayes's factor for each value of $p$, considering as the base model $p=2$.  Finally we provide our concluding remarks in Section \ref{con}.

\section{\texorpdfstring{$\alpha$}--attractor inflation and its generalizations}\label{generalizations}
In a recent article a new generalization of the basic $\alpha$-attractor model has been proposed~\cite{German:2021rin}. The basic model is defined by the potential~\cite{Kallosh:2013yoa}
\begin{equation}
V(\Psi) \propto \Psi^2 ,
\label{Ftanh} 
\end{equation}
where $\Psi \propto \tanh(\lambda \phi/M_{pl})$ makes $\phi$ a canonically normalized field identified with the inflaton in a class of phenomenological models of the type \cite{Akrami:2017cir}
\begin{equation}
\frac{1}{\sqrt{-g}}\mathcal{L}=\frac{1}{2}M_{pl}^2R-\frac{1}{2}M_{pl}^2\frac{(\partial_{\mu}\Psi)^2}{(1-\lambda^2\Psi^2)^2} -V(\Psi),
\label{Lagrangian} 
\end{equation}
where $M_{pl}$ is the reduced Planck mass given by $M_{pl}=2.44\times 10^{18} \,\mathrm{GeV}$. The potential \eqref{Ftanh} is naturally generalized to $V(\Psi) \propto \Psi^{p}$ giving rise to the form~\cite{Kallosh:2013yoa},
 \begin{equation}
V_t=V_0\tanh ^p\left(\lambda\frac{\phi}{M_{pl}}\right)\,.
\label{potanh} 
\end{equation}

On the other hand the generalization proposed in~\cite{German:2021rin} is given by $V(\Psi) \propto 1-(1-\Psi^2)^{p/2}$ resulting in the potential
\begin{equation}
V_s =V_0\left[1-\text{sech}^{p}\left(\lambda\frac{\phi}{M_{pl}}\right)\right]\,.
\label{potsech} 
\end{equation}

\begin{figure}[h!]
\begin{center}
\includegraphics[trim = 0mm  0mm 1mm 1mm, clip, width=10cm, height=7.cm]{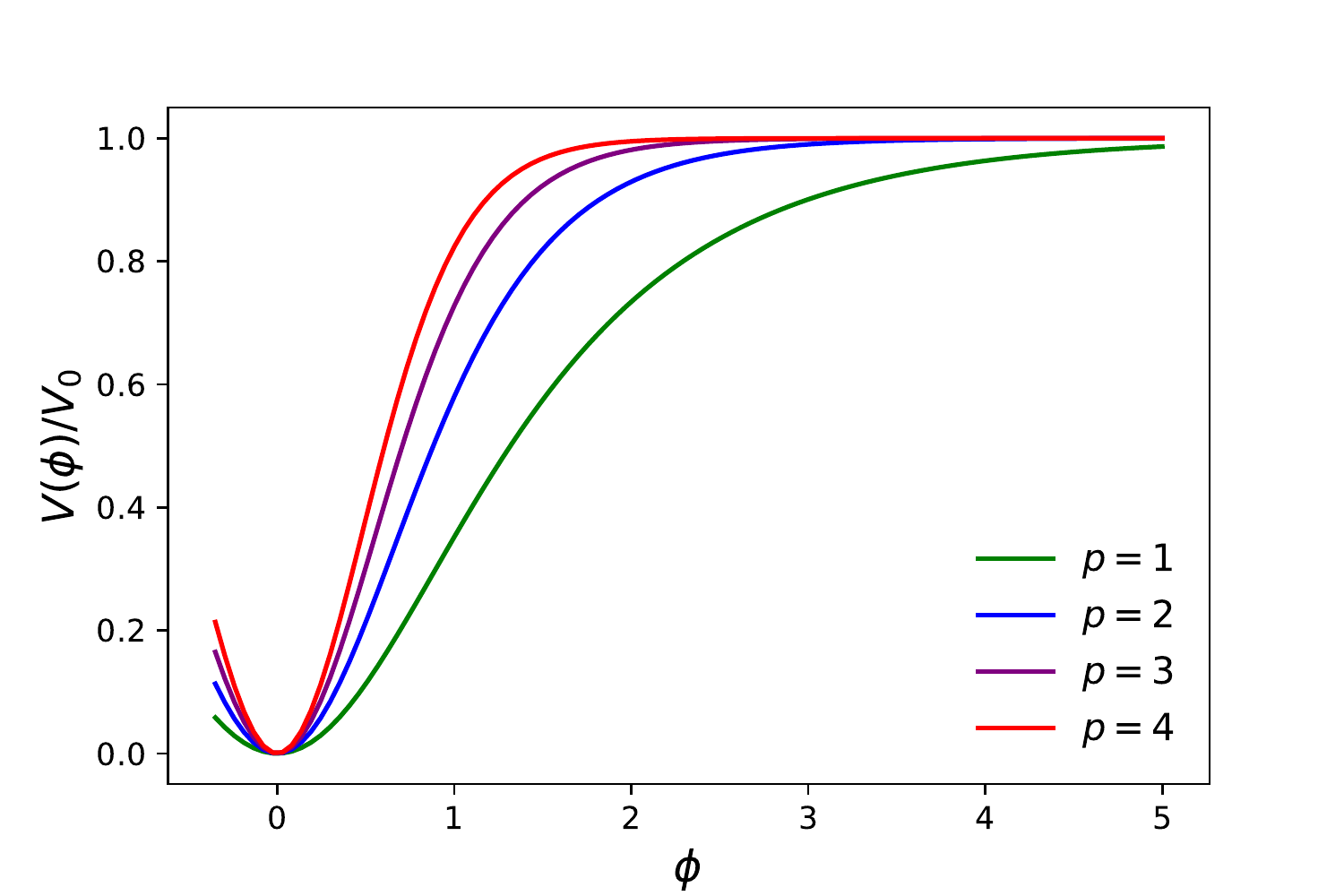}
\caption{\small Schematic plot of the $\text{sech}^p$ potential given Eq.~\eqref{potsech} for (from bottom to top) $p=1, 2, 3, 4$. The fact that the potential \eqref{potsech} is constructed on a positive definite function allows the possibility of having a well defined {\it inflationary} potential for practically any value of the parameter $p$ including odd and fractional values. }
\label{figpotsech}
\end{center}
\end{figure}

Notice that for $p=2$, the original potential \eqref{potanh} is recovered. The main distinction between \eqref{potanh} and \eqref{potsech} is that the latter involves a positive definite function, and  it thus produces a series of well-behaved {\it inflationary} potentials for practically any value of the parameter $p$ (including odd and fractional values of $p$). This is illustrated in Figure~\ref{figpotsech}.

Contrary to the potential~\eqref{potanh}, the generalized potential~\eqref{potsech} also has the peculiarity that the first term in the expansion around the origin is quadratic with the parameter $p$ appearing only as a multiplicative factor of $\phi^2$ and of higher order terms as:
\begin{equation}
\frac{V_s}{V_0} =\frac{1}{2}p \left(\lambda \frac{\phi}{M_{pl}}\right)^2-\frac{1}{24}p(2+3p)\left(\lambda \frac{\phi}{M_{pl}}\right)^4+\cdot\cdot\cdot \, .
\label{potexpanded}
\end{equation}

\noindent The properties of this potential and its advantages have been already analyzed in detail~\cite{German:2021rin}. In the present work, our main goal is to fully explore the scope of such inflationary model, making use of both numerical tools and statistical methods, in order to assess its viability and evaluate specific predictions on physical observables.

\subsection{\texorpdfstring{$\alpha$}--attractor inflation in \textsc{class}}

To simulate the evolution of linear perturbations in the universe and to compute the CMB anisotropies as well as large scale structure observables, we use the numerical code \textsc{class}~\cite{Lesgourgues:2011re}. The usual expressions for the slow-roll parameters $\eta$ and $\epsilon$ are defined in the \texttt{primordial.c} module as follows:
\begin{equation}
    \eta =M_{pl}^2\frac{V^{\prime\prime}}{V}\, ,\quad
\epsilon = \frac{M_{pl}^2}{2} \left( \frac{V^{\prime}}{V} \right)^2\, ,
\end{equation}
while the observables corresponding to the scalar spectral index $n_s$, the tensor-to-scalar ratio $r$, and the scalar and tensor  amplitudes at horizon crossing $A_s$ and $A_t$ are, respectively, given by
\begin{equation}
    n_s =1+2\eta-6\epsilon \, ,\quad
r= 16\epsilon\, ,\quad
A_s = \frac{1}{24\pi^2 \epsilon}\frac{V}{M_{pl}^4}\, ,\quad
A_t = \frac{2}{3\pi^2}\frac{V}{M_{pl}^4}.
\end{equation}
where $M_{pl}\equiv 1/\sqrt{8\pi G_N}$ is the reduced Planck mass which we take as $M_{pl}= 1/\sqrt{8\pi}.$ For the specific model Eq.~\eqref{potsech} the amplitudes at horizon crossing are given by
\begin{eqnarray}
A_s &=& \frac{V_0}{M_{pl}^4}\frac{\left[\text{sech}\left(\lambda\frac{\phi_{\star}}{M_{pl}}\right)\right]^{-2p}\left[1-\text{sech}^{p}\left(\lambda\frac{\phi_{\star}}{M_{pl}}\right)\right]^3}{12p^2\pi^2\lambda^2\left[1-\text{sech}^{2}\left(\lambda\frac{\phi_{\star}}{M_{pl}}\right)\right]}\, ,\label{A_s_class} \\
A_t &=& \frac{2 V_0}{3\pi^2 M_{pl}^4}\left[1-\text{sech}^{p}\left(\lambda\frac{\phi_{\star}}{M_{pl}}\right)\right]\, ,
\end{eqnarray}
where $\phi_{\star}$ is the inflaton value at the time scales leave the horizon. Finally, the correspondence between the parameters of the inflationary model ($V_0\, , \lambda\, , p$) with those in \textsc{class} (\texttt{Vparam0, Vparam1, Vparam2}) is
\begin{equation}
    \texttt{Vparam0}=\frac{V_0}{M_{pl}^4}\, ,\quad
    \texttt{Vparam1}=\lambda\, ,\quad
    \texttt{Vparam2}=p\, .
\end{equation}

\subsection{\texorpdfstring{$\alpha$}--attractor inflation: particular cases}\label{semianalytical}

The cases $p=1, 2$ and $4$ have been studied analytically in some detail in ref.~\cite{German:2021rin}. Due to the complexity of the equations, only in the $p=2$ case has it been possible to find an analytical expression in closed form for the tensor-to-scalar ratio $r$ in terms of the scalar spectral index $n_s$ and the number of $e$-folds $ N_{\star}$ during inflation. This is given by
\begin{equation}
r = \frac{4(N_{\star}\delta_{n_s} -2)^2}{1+N_{\star} (N_{\star} \delta_{n_s}-2)}\;,
\label{rdensNkep2}
\end{equation}
where $\delta_{n_s}$ is defined as $\delta_{n_s}\equiv 1-n_s$. It is not hard to show that $n_s$ is bounded by~\cite{German:2021rin}
\begin{equation}
1-\frac{2}{N_{\star}} \leq n_s < 1-\frac{2}{N_{\star}}+\frac{1}{N_{\star}^2}\;.
\label{conditions}
\end{equation}

Thus, we see that the range of $n_s$ is very small and proportional to $1/N_{\star}^2$. These models have been further studied in ref.~\cite{German:2020cbw}, where constraints from the reheating epoch are imposed. 

In addition to the cases previously studied, here we will add the case $p=3$, for which it is possible to write the parameters, $V_0$ and $\lambda$, in terms of the observables $n_s$ and $r$ and use the ranges reported by Planck and other experiments to constrain their ranges. The resulting expressions, however, are too intricate, and not worth writing explicitly.

Note that Eq. (2.12) implies that the range of allowed values for $n_s$ is dictated by that of $1/N_{\star}^2$. In the considered model, $\eta$ is of order $1/N_{\star}$, so it is of order $\eta^2$. This shows that, for example, in equation (2.7), contributions of order $\eta^2$ must be included in $n_s$. However, the above has been written in the first order Slow-Roll approximation for illustrative purposes, since the Bayesian analysis is performed numerically without approximations.

\section{Numerical Analysis}\label{numerical}

\begin{figure}[h!]
\centering 
\includegraphics[width=0.7\textwidth]{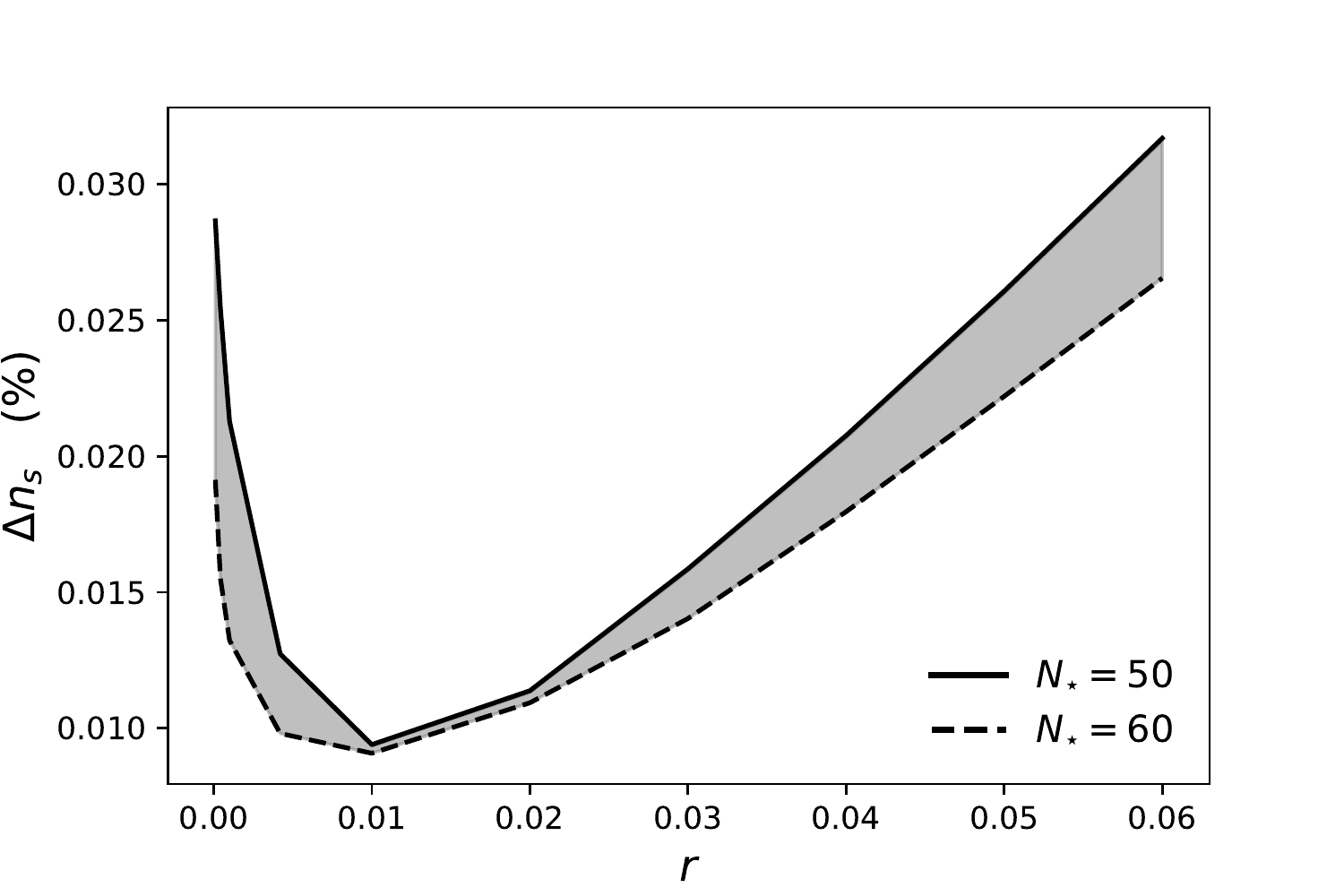}
\caption{\label{r_ns_comp} Relative differences in the value of $n_s$ between the semi-analytical and numerical results for a given $r$. The gray band covers the cases from $p=1$ to $p=4$. See text for more details.}
\end{figure}

As mentioned in Section~\ref{semianalytical}, it has been possible to obtain analytical expressions for the case $p=2$, whereas for $p=1$ and $p=4$, semi-analytical formulas have been useful to infer some of the properties of such models (see~\cite{German:2021rin}). As a first step in the analysis of these models,  we implement the $\alpha$-attractor model in \textsc{class}, and we solve numerically the inflationary phase considering the inflaton potential given by~\eqref{potsech}. In particular, we take as reference point the study in ref.~\cite{German:2021rin}, and consider at first the same values of the exponential $p=1\, , 2\, , 4$ (for the complete analysis, however, we include in the following section the odd case $p=3$). For the same exact value obtained from the semi-analytical solutions for the tensor-to-scalar ratio $r$, the $e$-fold number $N_{\star}$, and the potential parameters $V_0$ and $\lambda$, we have obtained the numerical values of the spectral index $n_s$ from \textsc{class}.
This can be seen in Figure~\ref{r_ns_comp}, where the relative difference for the spectral index $\Delta n_s$ between the semi-analytical and numerical results, is shown. The gray band covers the region for each value of $r$ and for each case $p=1,2,4$, of the percentage difference on $n_s$ on the range of $N_{\star} \in [50\, , 60]$. In particular, the solid line corresponds to the highest difference which occurs for $p=1$ and $N_{\star}=50$, whereas the lower difference (dashed line) is obtained for the case $p=4$ and $N_{\star}=60$. The model under study has the particularity that an expansion around its minimum always has a quadratic term as its main contribution, the parameter $p$ playing a weak role not as a power of $\phi$ as in the $\tanh^p$ model but as a coefficient of the powers of $\phi$ only (see~\eqref{potexpanded}). Therefore, we can think that perturbative reheating occurs for $\omega_{re}\approx 0$. In this case and following \cite{Martin:2010kz}, \cite{Dai:2014jja} the priors for $N_{\star}$ have been calculated in~\cite{German:2020cbw}, where the inflationary constraints coming from reheating are studied and given there in Table 1. Thus, the priors we use for $N_{\star}$ are consistent with a reheating epoch.

 Figure~\ref{r_ns_comp} thus shows that the values for $n_s$ are recovered in a very good approximation. In fact, for all cases the relative differences are $\lesssim 0.030\%$. 
 The numerical solutions obtained from \textsc{class} for the scalar amplitude $\ln(10^{10}A_s^{{\rm{\textsc{class}}}})=3.045$, were always consistent with the current value reported by Planck Collaboration 2018~\cite{Planck:2018jri}: $\ln(10^{10}A_s)=3.044\pm0.014$. This is important since, with no \textit{a priori} constraints for the range of values of the potential parameters $V_0$ and $\lambda$, solving the inflationary phase with \textsc{class} allowed us to determine the values of potential~\eqref{potsech} consistent with the current observed values of $n_s$ and $A_s$, when $p$ and $N_{\star}$ are specified. This is useful to establish the priors for the parameters of the $\alpha$-attractor potential.

\begin{figure}[h!]
\centering 
\includegraphics[width=0.66\textwidth]{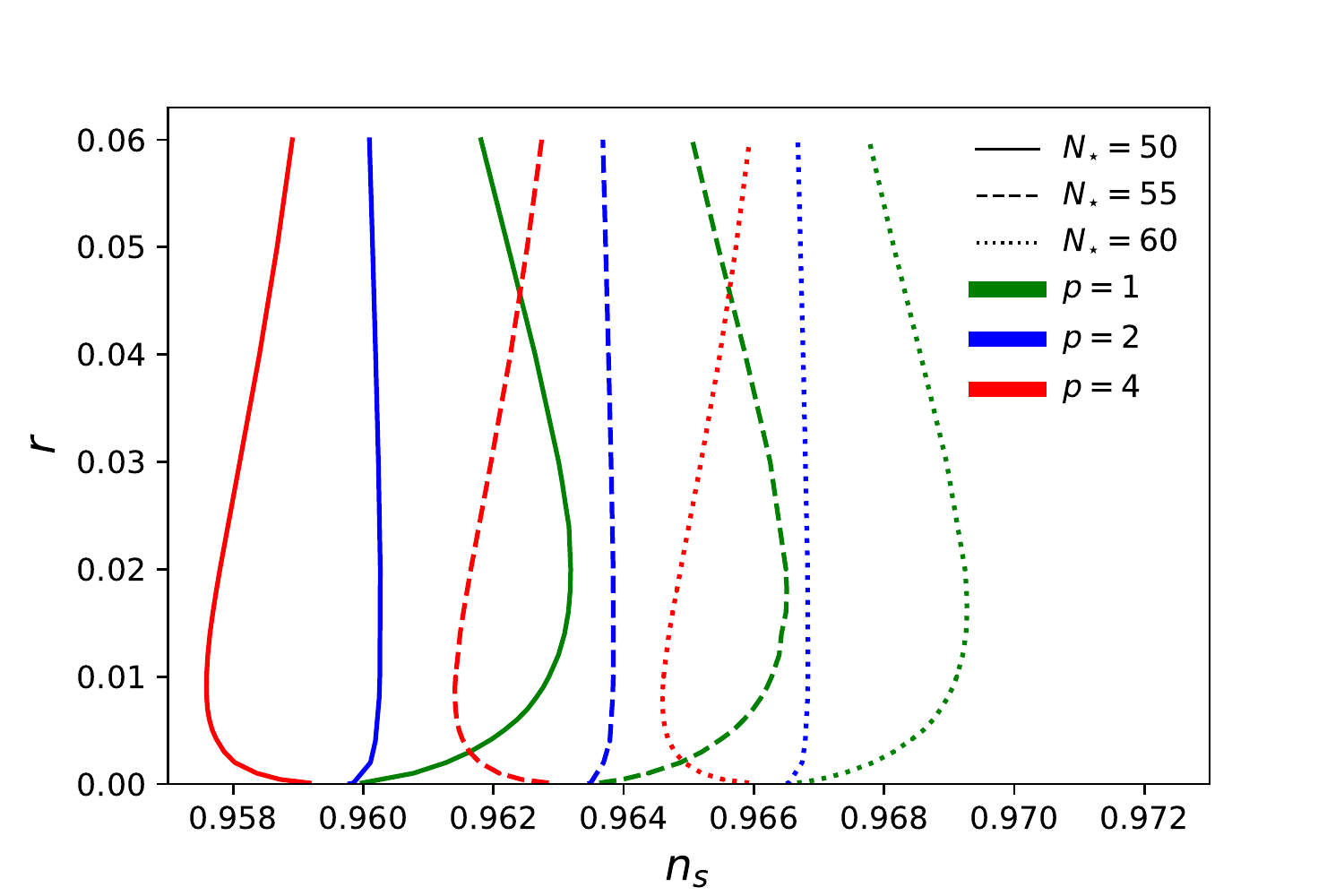}
\caption{\label{r_ns_all} $(n_s-r)$-plane obtained for the potential~\eqref{potsech} with $p=1,2,4$ and $N_{\star}=50\, , 55\, , 60$. See text for more details.}
\end{figure}

Additionally, in Figure~\ref{r_ns_all} we observe that as the number of e-fold increases, the value of $n_s$ also increases for a given $p$. On the other hand, we can see that the value of $r$ is not restricted for any value of $N_{\star} = 50\, , 55\, , 60$ (solid, dashed, and dotted line respectively), or by any of the exponents $p=1\, , 2\, , 4$ (green, blue, and red respectively). Both aspects mentioned above, i.e., the correlation between $N_{\star}$ and $n_s$, and the lack of a preferred value of $r$, will be looked at in detail through the statistical analysis of  Section~\ref{statistical}.

\begin{figure}[h]
\centering 
\includegraphics[width=0.7\textwidth]{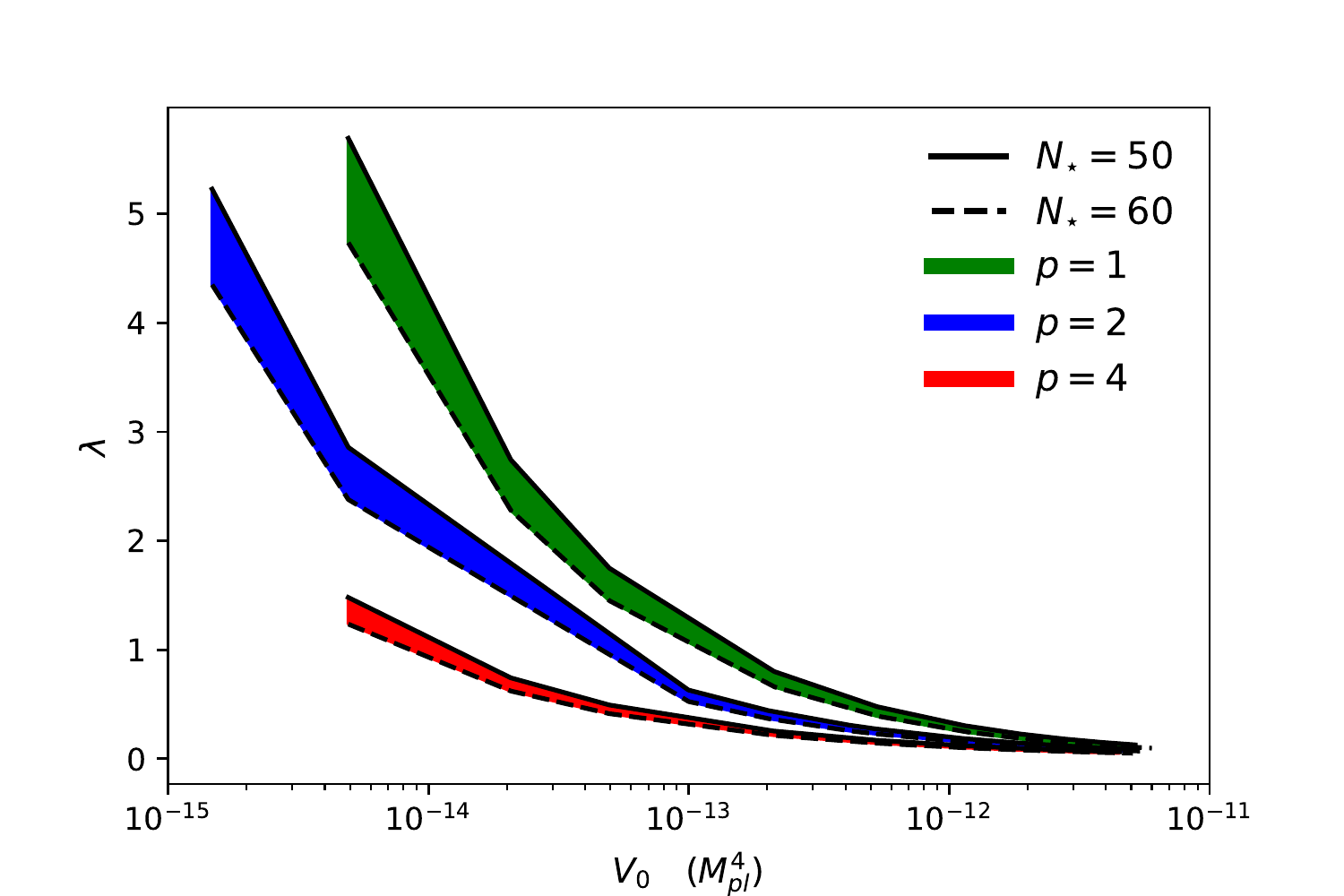}
\caption{\label{range_V0_lambda} Range of values for $V_0$ and $\lambda$ for a given $p$ and $N_{\star}$. For low values of $\lambda$, all the curves converge. See text for more details.}
\end{figure}

The range of values for the potential parameters $V_0$ and $\lambda$, determined by the corresponding range of observables obtained from \textsc{class}, is illustrated, for the relevant cases, in Figure~\ref{range_V0_lambda}.  It is important to mention that the numerical precision allows for values of the tensor to scalar ratio as small as $r\sim 10^{-5}$, which in turn allows for larger values of the potential parameter $\lambda$ for each case of $p$ considered, given $N_{\star}$. Therefore, our priors on $\lambda$ will be numerically bounded from above, but for all the cases such maximum values of $\lambda$ will be consistent with the current constraints on $r<0.032$~\cite{Tristram:2021tvh}.

\section{Statistical analysis}\label{statistical}


Following the approach of Section~\ref{numerical}, we have included in our analysis a new the case for the potential~\eqref{potsech}: $p=3$, that is, a new, odd value of $p$. In order to obtain the priors for $[V_0\, , \lambda]$ when $p=3$, we have made an interpolation between the range of values obtained for such parameters when considering $p=2$ and $p=4$. Moreover, we have verified that the values obtained for the cosmological parameters $A_s\, , n_s$ and $r$, where consistent with the reported values by observations, just as we did for the cases $p=1\, , 2\,\mathrm{and}\, 4$ in the previous Section.

According to Bayes' theorem, the probability of a model $M$ with a set of parameters $\Theta$, in light of the observed data $D$, is given by the  \textit{Posterior} $\mathcal{P}$: 
\begin{equation}
    \mathcal{P}(\Theta \mid D,M)= \frac{\mathcal{L}(D\mid \Theta,M)\Pi(\Theta\mid M)}{\mathcal{E}(D\mid M)},
    \label{eq:bayes}
\end{equation}

\noindent where $\mathcal{L}$ is the Likelihood function, $\Pi$ represents the set of  Priors, containing the \textit{a priori} information about the parameters of the model and $\mathcal{E}$ is the  Evidence. For any value of $p$, the parameter space we are considering is given by $[100\omega_b\, , \omega_{cdm}\, , 100 \theta_s\, , \tau_{reio}\, , \log V_0\, , \lambda\, , N_{\star}]$, which are the standard cosmological parameters for the amount of baryons $\omega_b$ and cold dark matter $\omega_{cdm}$, the angular diameter distance $\theta_s$, the optical depth $\tau_{reio}$, the potential parameters $V_0\, , \lambda$, as well as the number of e-fold $N_{\star}$. The priors are indicated in Table~\ref{priors}. On the other hand, as a derived parameters of interest we will obtain posteriors for $n_s$ and $r$ as well.
\begin{table}[h!]
\centering
\begin{tabular}{cccc}
\hline
\hline
parameter & mean & prior min & prior max      \\
\hline
\hline
$100\omega_b$ & $2.2377$ & $2.2$ & $2.6$      \\
$\omega_{cdm}$ & $0.12010$ & $0.10$ & $0.13$            \\
$100 \theta_s$ & $1.04110$ & $1.0$ & $1.1$       \\
$\tau_{reio}$ & $0.0543$ & $0.004$ & $0.07$         \\
$\log V_0$ & $-13$ & $-14$ & $-12$         \\
$\lambda$ & $2$ & $0$ & $4$         \\
$N_{\star}$ & $55$ & $50$ & $60$         \\
\hline
\end{tabular}
\caption{Priors for cosmological parameters, including those from the inflationary potential, i.e., $V_0$ and $\lambda$. See text for more details.}
\label{priors}
\end{table}

For a given model $M$, the Bayesian Evidence $\mathcal{E}$ (hereafter simply the evidence) is the normalizing constant in the right hand side of Eq. (\ref{eq:bayes}). It normalises the area under the posterior $\mathcal{P}$ to unity, and is given by
\begin{equation}
  \mathcal{E}(D\mid M)= \int d\Theta\, \mathcal{L}(D|\Theta,M)\Pi(\Theta|M)\, . 
  \label{eq:evidence}
\end{equation}

The evidence can be neglected in model fitting, but it becomes important in model comparison. In fact, when comparing two different models $M_1$ and $M_2$ using Bayes' theorem \eqref{eq:bayes}, the ratio of posterior probabilities of the two models $\mathcal{P}_1$ and $\mathcal{P}_2$ will be proportional to the ratio of their evidences, that is
\begin{equation}
    \frac{\mathcal{P}_1(\Theta_1 \mid D,M_1)}{\mathcal{P}_2(\Theta_2 \mid D,M_2)} = \frac{\Pi_1(\Theta_1|M_1)}{\Pi_2(\Theta_2|M_2)}\frac{\mathcal{E}_1(D\mid M_1)}{\mathcal{E}_2(D\mid M_2)}\, .
\end{equation}

This ratio between posteriors leads to the definition of the \textit{Bayes Factor} $B_{12}$, which in logarithmic scale is written as
\begin{equation}
    \log B_{12} \equiv \log \left[\frac{\mathcal{E}_1(D\mid M_1)}{\mathcal{E}_2(D\mid M_2)}\right] = \log \left[\mathcal{E}_1(D\mid M_1)\right] - \log \left[\mathcal{E}_2(D\mid M_2)\right] \, .
    \label{BF}
\end{equation}

If $\log B_{12}$ is larger (smaller) than unity, the data favours model $M_1$ ($M_2$). To assess the strength of the evidence contained in the data, Jeffreys \cite{Jeffreys1961} introduces an empirical scale, see Table~\ref{Jeffreyscale}. 
\begin{table}[h!]
\centering
\begin{tabular}{cc}
\hline
\hline
$2\log{B_{12}}$    & Strength  \\
\hline
\hline
$<$ 0      & Negative (support $M_2$)      \\
0 - 2.2 & Weak       \\
2.2 - 6 & Positive     \\
6 - 10      & Strong     \\
$>10$ & Very strong \\
\hline
\end{tabular}
\caption{Jeffreys' scale to quantify the strength of evidence for a corresponding range of the Bayes factor. We follow the convention of \cite{Kass:1995loi,10.2307/2337598} in presenting a factor of two with the natural logarithm of the Bayes factor.}
\label{Jeffreyscale}
\end{table}

\begin{figure}[h!]
\centering 
\includegraphics[width=0.48\textwidth]{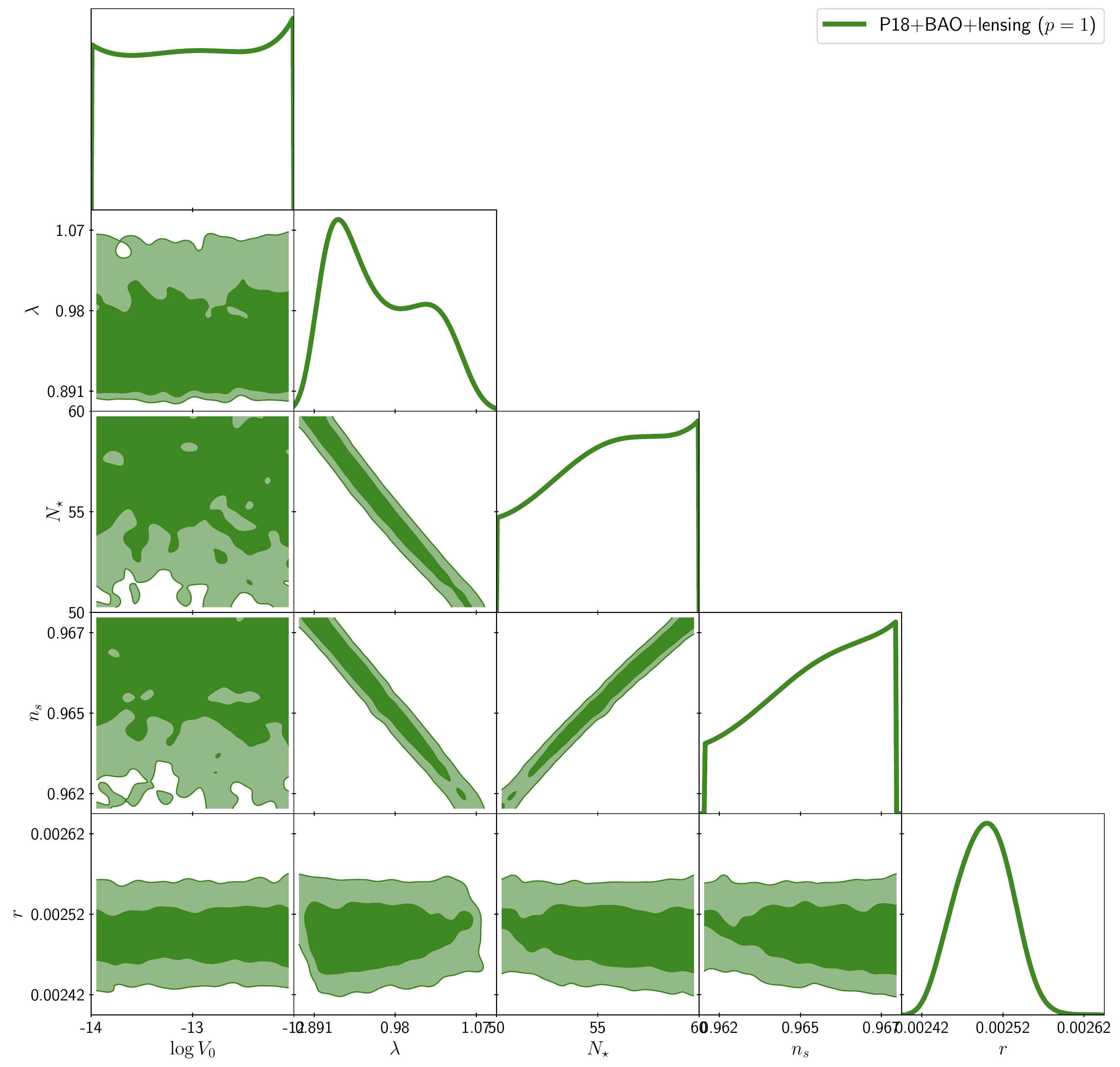}
\includegraphics[width=0.48\textwidth]{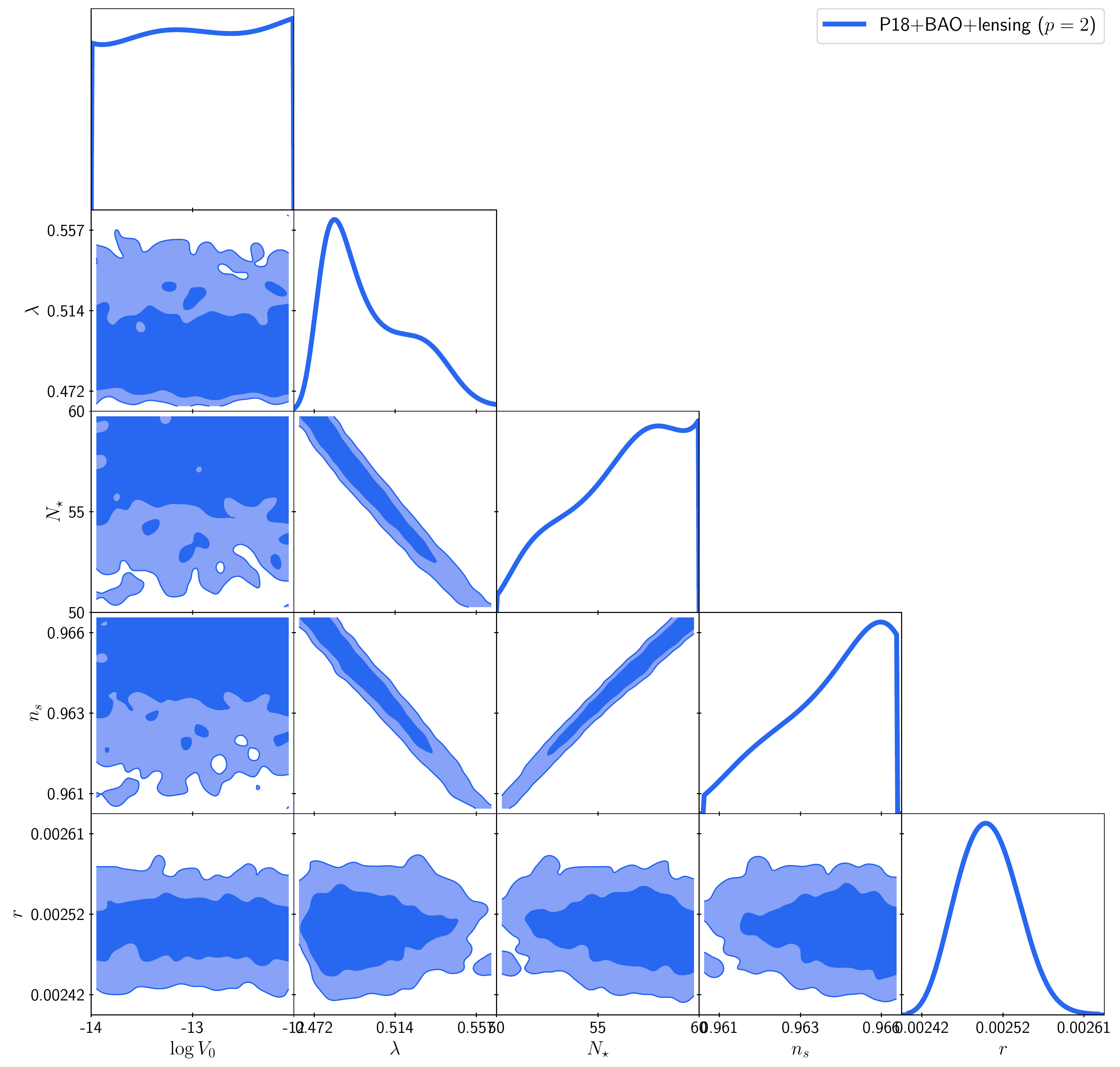}
\includegraphics[width=0.48\textwidth]{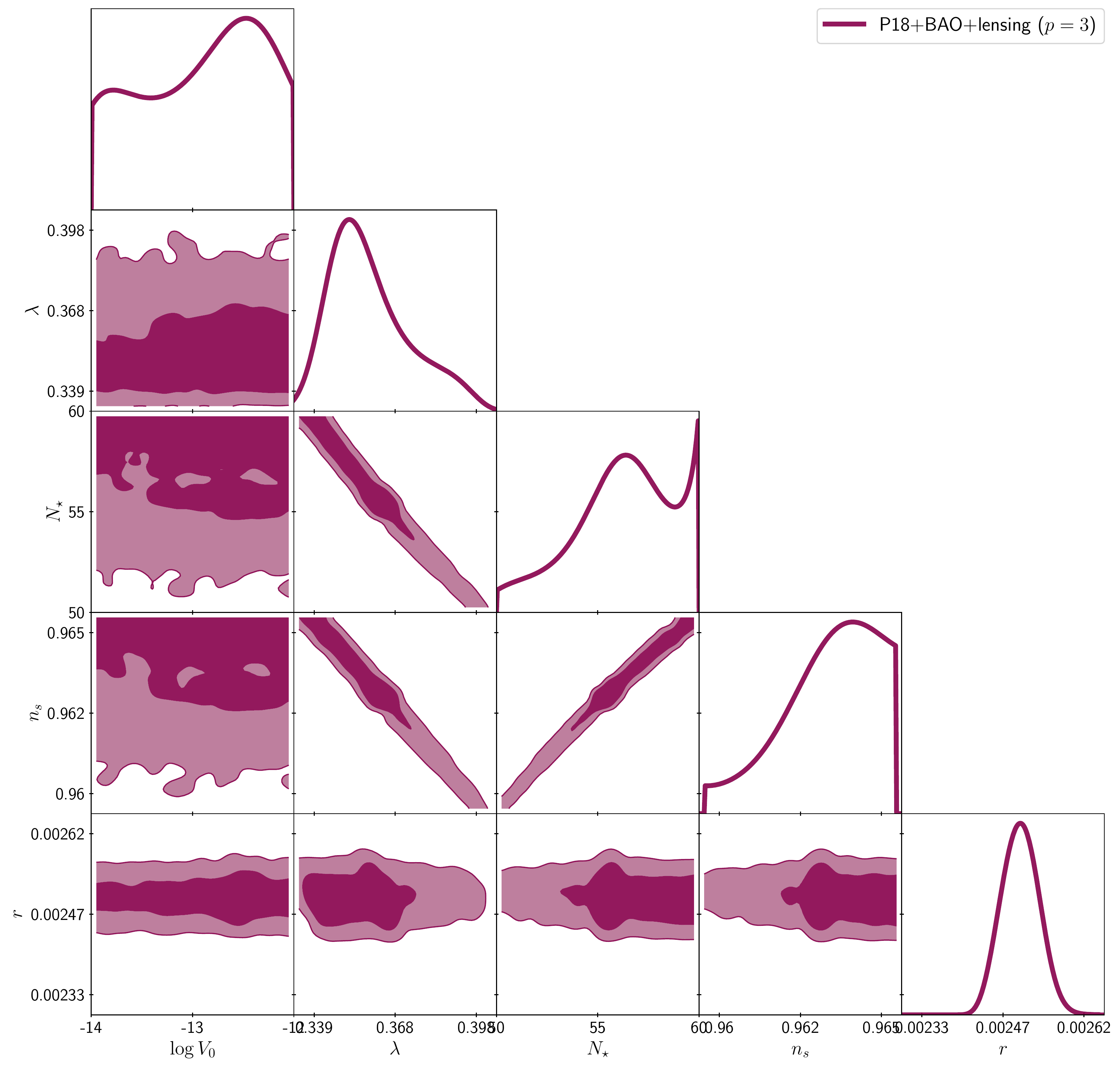}
\includegraphics[width=0.48\textwidth]{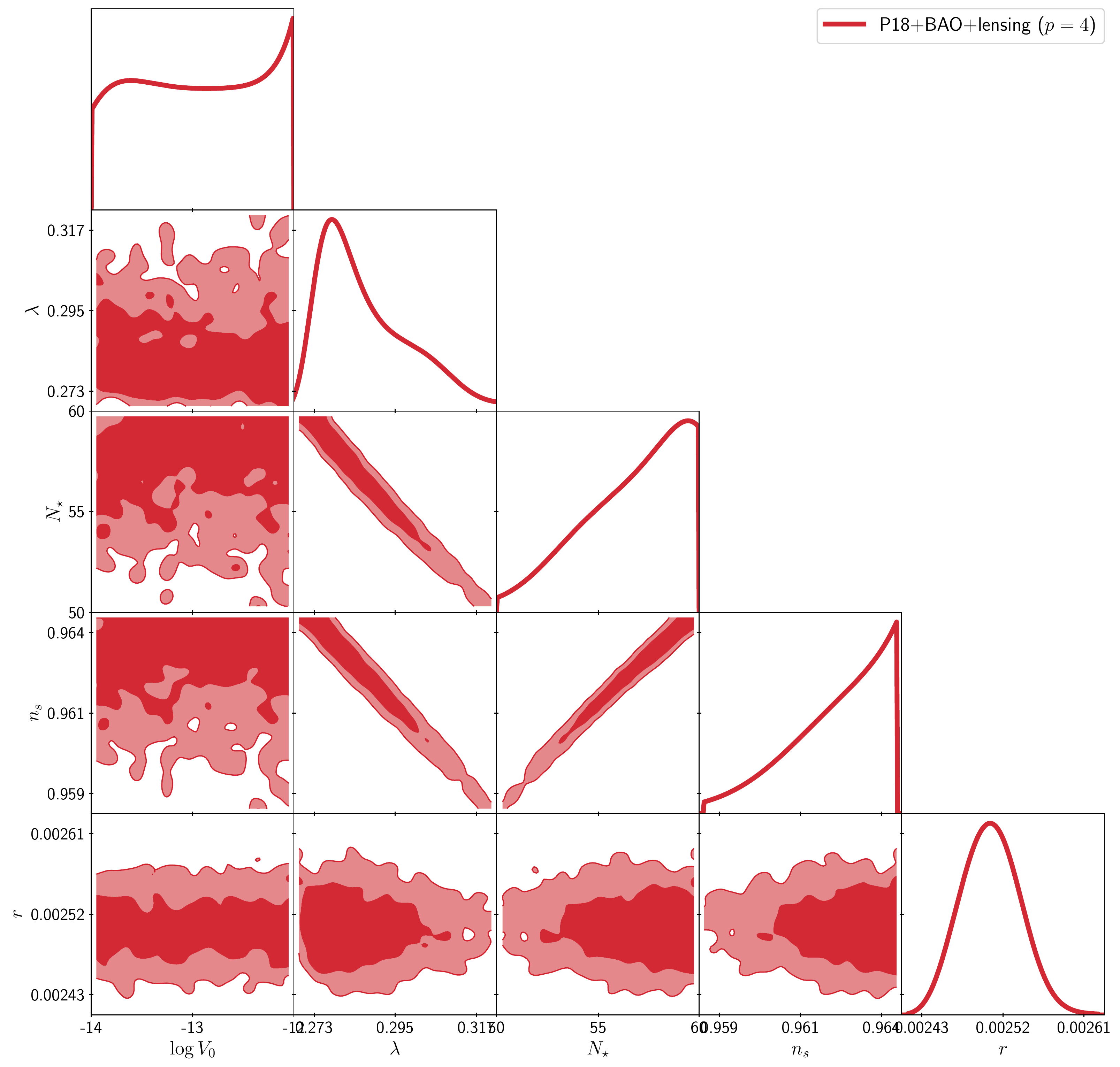}
\caption{\label{post} 1D and 2D posteriors for the $\alpha$-attractor potentials with $p=1$ (top left), $p=2$ (top right), $p=3$ (bottom left), and $p=4$ (bottom right). We focus particularly in the following parameters: $\left[\log V_0\, , \lambda\, , N_{\star}\, , n_s\, , r\right]$. See text for more details.}
\end{figure}

We show in Figure~\ref{post} the posteriors for the $\alpha$-attractor parameters $\log V_0\, ,$ and $\lambda$, as well as for the standard cosmological inflation parameters $N_{\star}\, , n_s\, ,$ and $r$. The posteriors were obtained by using the CMB likelihood from Planck Satellite 2018, which is publicly available in \url{http://pla.esac.esa.int/pla/\#cosmology}. Thus, we generated the corresponding Markov Chain Monte Carlo (MCMC) with \textsc{Monte Python}\footnote{Notice that we do not include the posteriors for the standard cosmological parameters $100\omega_b\, , \omega_{cdm}\, , 100 \theta_s\, , \tau_{reio}$ in Figure~\ref{post}, since they do not present major changes with respect to their tightly constrained values from Planck~\cite{Planck:2018vyg}. For such parameters we obtained (specifically for $p = 3$): $100\omega_b = 2.24^{+0.0127}_{-0.0121}\, , \omega_{cdm}=0.12^{+0.000786}_{-0.000774}\, , 100 \theta_s = 1.04^{+0.000269}_{-0.000285}\, , \tau_{reio} = 0.0534^{+0.00717}_{-0.00606}\, .$}. The (logarithmic) amplitude of the $\alpha$-attractor potential presents a nearly flat posterior for all values of the exponential $p$, indicating that $V_0$ has approximately the same probability to match CMB observations for values in the range $V_0 = \left[10^{-14}\, , 10^{-12}\right]$. On the other hand, we show for the first time, constraints on the theoretical parameter encoding the curvature of the Kähler manifold, $\lambda$, as displayed in Table~\ref{lambda_values}:
\begin{table}[h!]
\centering
\begin{tabular}{ccccc}
\hline
\hline
$p$    & 1 & 2 & 3 & 4  \\
\hline
\hline\\[-0.3cm]
$\lambda$ & $0.962^{+0.0068}_{-0.0144}$ & $0.503^{+0.0027}_{-0.0065}$ &  $0.361^{+0.0003}_{-0.0001}$ & $0.287^{+0.0012}_{-0.0032}$   \\[0.1cm]
\hline
\end{tabular}
\caption{Mean and standard deviation for inflaton potential parameter $\lambda$ for each value of $p$.}
\label{lambda_values}
\end{table}

Additionally, we have obtained that any value of $p$ study here, the tensor to scalar ratio is consistent with
\begin{equation}
    r\simeq 2.5\times 10^{-3} \pm 10^{-5}\, ,
    \label{pred_r}
\end{equation}
as can be seen in Figure~\ref{post}. This constitutes a prediction of the family of $\alpha$-attractor models with potential~\eqref{potsech}. Thus, as opposed to the semi-analytical results obtained in Section~\ref{numerical}, the statistical analysis of data constrains the value of $r$. We find interesting that all the models considered in this work agree in one specific value for the tensor-to-scalar ratio.

From the posteriors we also confirm the relation we obtained numerically for $N_{\star}$ and $n_s$: as $N_{\star}$ increases, $n_s$ increases as well (see disscusion in Section~\ref{numerical} and Figure~\ref{r_ns_all}), that is, these quantities are correlated. When broadening the priors for these parameters we found a long delay in convergence and practically no change in this behavior. On the other hand, it is also evident from the posteriors that $\lambda$ is anti-correlated with either $N_{\star}$ or $n_s$. This implies lower values of $\lambda$ to be preferred by CMB observations according to the estimated value of $N_{\star}$, and the observed value of $n_s$. In Section~\ref{con} we discuss the corresponding values for the parameter in the original notation $\alpha$, and the physical implications in field space.

With the posteriors at hand, we are en route to calculate the evidence. This can be computed through a variety of techniques~\cite{de1958asymptotic,tierney1986accurate,bleistein1986asymptotic,mackay1997ensemble,vsmidl2006variational,Skilling:2006gxv,2008arXiv0801.3887C,2009AIPC.1193..251R}, and several programs have been developed to compute the evidence for cosmological models by solving the multidimensional integral given by Eq.~\eqref{eq:evidence}~\cite{Feroz:2007kg,Feroz:2008xx,Feroz:2013hea,Handley:2015fda,2015MNRAS.453.4384H} (for a Bayesian analysis of several inflationary models, some of the authors have used these numerical tools, see for instance~\cite{Barbosa-Cendejas:2017pbo}).

One alternative to calculate the evidence is given by~\cite{Heavens:2017afc}, where the unnormalized posterior $\tilde{\mathcal{P}}(\Theta\mid D,M)$ is proportional to the number density $n(\Theta\mid D,M)$, we have, $\tilde{\mathcal{P}}(\Theta\mid D,M) = a\ n(\Theta\mid D,M)\,$. Since the number density is given by
\begin{equation}
    n(\Theta\mid D,M) = N\ \mathcal{P}(\Theta\mid D,M) = N\ \frac{\tilde{\mathcal{P}}(\Theta\mid D,M)}{\mathcal{E}(D \mid M)},
\end{equation}

\noindent where $N$ is the lenght of the chain, then,
\begin{equation}
    \quad \mathcal{E}(D \mid M) = a\ N\, ,
\end{equation}
 
\noindent and thus, once knowing the proportionality constant $a$, the evidence $\mathcal{E}$ can be directly obtained from the MCMC chains. The open source software \textsc{MCEvidence}~\cite{Heavens:2017afc} is designed to compute the Bayesian evidence from MCMC sampled posterior distributions. The code takes the $k-$th nearest-neighbor distances in parameter space using the Mahalanobis distance to estimate the Bayesian evidence from the MCMC samples provided by the chains.

Employing this approach, Table~\ref{Jeffrey_alpha} shows the (logarithmic) evidence $\ln\mathcal{E}_p$, the Bayes factor $2\log{B_{p2}}$ when choosing the case $p=2$ as the reference model, and the strength of the evidence for each value of $p$. The evidence for the case $p=1$ was already studied in~\cite{Martin:2013tda,Martin:2013nzq} (Mutated Hilltop Inflation (MHI) in the notation of the mentioned references), but in such study it was compared to the Higgs Inflation (HI) potential as the base model. In that case, the authors analyzed three separate cases for the prior of the parameter $\mu$ which plays the role of our $1/\lambda$ within the hyperbolic secant function. They found that the MHI is disfavoured over HI, just as in our case when comparing $p=1$ with the base model we consider, $p=2$. 
\begin{table}[h!]
\centering
\begin{tabular}{cccc}
\hline
\hline
$p$ &  $\ln\mathcal{E}_p$ & $2\log{B_{p2}}$    & Strength  \\
\hline
\hline
$1$ & $-1414.99$ & $-2.94$  & Negative (support $p=2$)      \\
$2$ & $-1413.52$ & $0$      & Reference model      \\
$3$ & $-1415.46$ & $-3.88$  & Negative (support $p=2$)         \\
$4$ & $-1410.60$ & $5.84$   & Positive (support $p=4$)      \\
\hline
\end{tabular}
\caption{Jeffreys' scale for each of the $\alpha$-attractor models, according to table~\eqref{Jeffreyscale}.}
\label{Jeffrey_alpha}
\end{table}

\begin{figure}[h!]
\centering 
\includegraphics[width=0.67\textwidth]{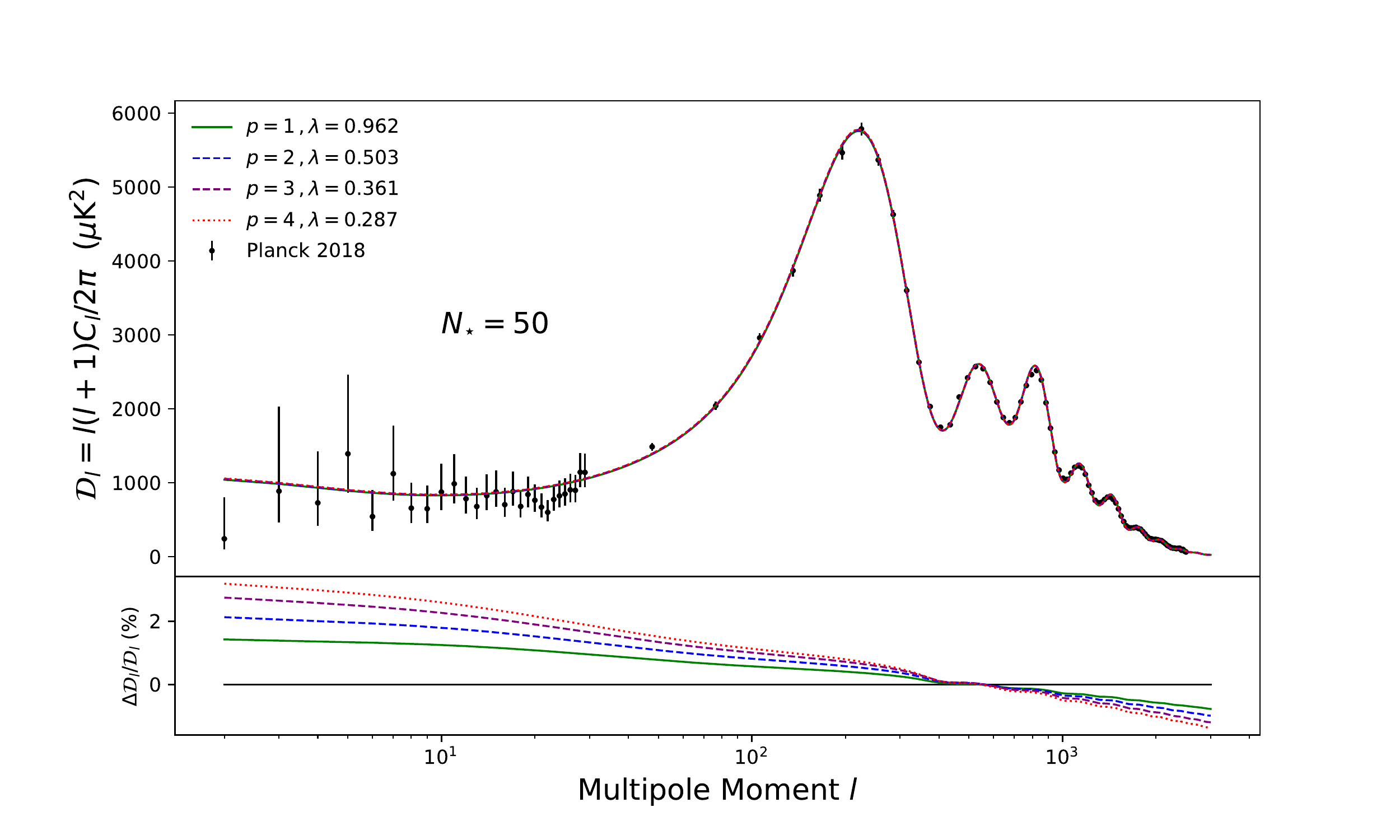}
\includegraphics[width=0.67\textwidth]{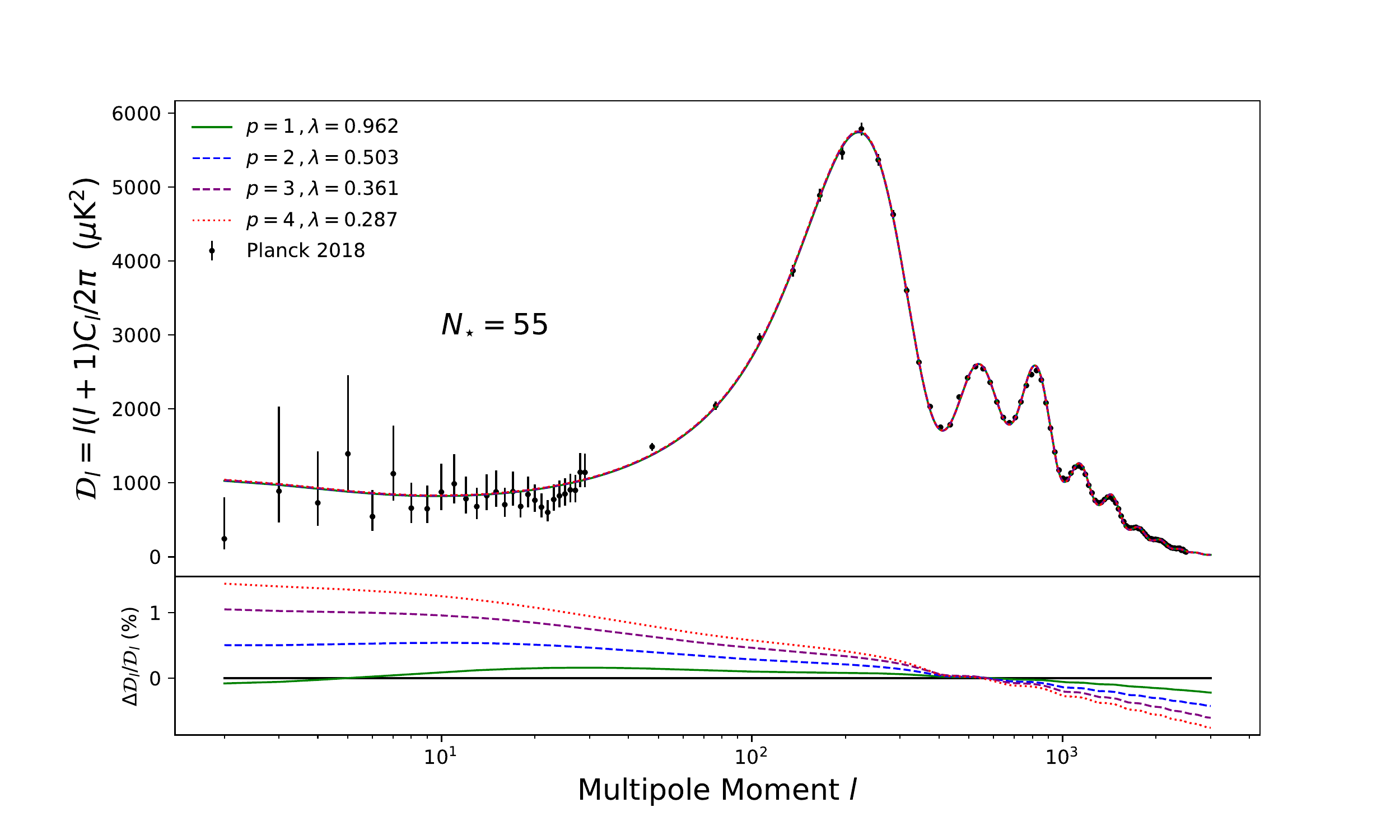}
\includegraphics[width=0.67\textwidth]{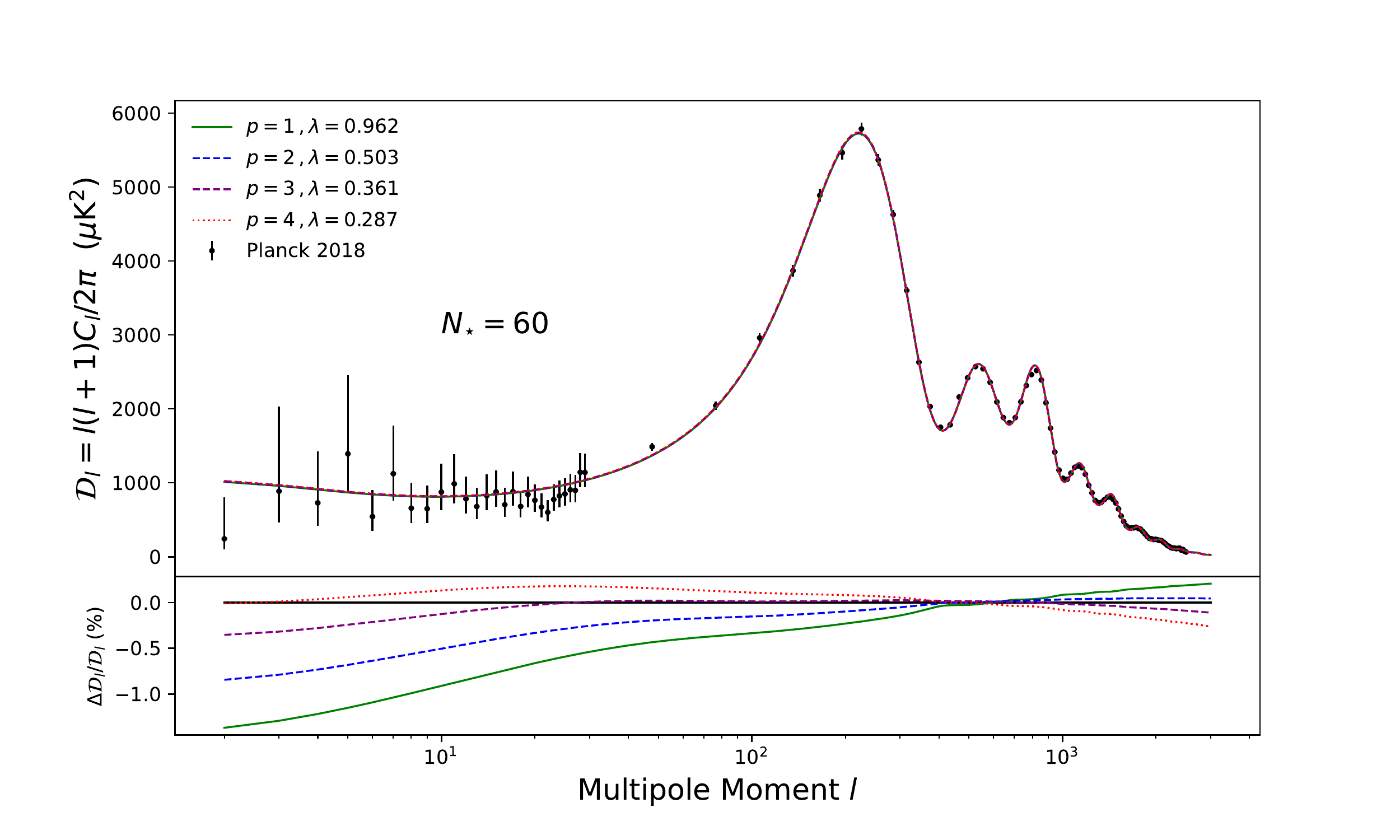}
\caption{\label{cmb_ani} CMB anisotropies (upper panels of each plot) and their relative difference (lower panels) for each value of $p$, with $\lambda$ given by the mean value inferred in the statistical analysis in Section~\ref{statistical} (see Table~\ref{lambda_values}). It can be seen that, as $N_{\star}$ increases from $50\, , 55\, , 60$ (top, middle, and bottom respectively), the relative differences decreases when $p=4$. The Planck 2018 data from {\url{http://pla.esac.esa.int/pla/\#cosmology}}.}
\end{figure}

Thus, whereas odd values of $p$ are not preferred over the base model with $p=2$, the case $p=4$ results favoured by CMB observations with a \textit{positive} strength according to the Jeffreys' scale. We can understand why this is so by looking at the spectrum of CMB anisotropies shown in Figure~\ref{cmb_ani}. The figure displays the spectrum of the CMB radiation taken by the Planck Satellite Collaboration~\cite{Planck:2018vyg}, together with the numerical solutions for each value of $p=1\, , 2\, , 3\, , 4$. As stated in Section~\ref{statistical}, $V_0$ can take any value within the priors. Thus, we take values of $V_0 \simeq 10^{-13}$, which are consistent with the observed $A_s\, , n_s$ and $r$\footnote{For this last parameter we obtained the numerical value matching our result $r\simeq 0.0025$.}. In the case of $\lambda$, we have used the mean value reported in Table~\ref{lambda_values}. The remaining cosmological parameter to consider is $N_{\star}$. From the posteriors we see that larger values of $N_{\star}$ are most likely, according to CMB data. Consequently, the numerical solutions will better fit the data as the $e$-fold number gets larger. This is precisely the case: for $N_{\star}=50$ and $55$ (top and middle plots respectively in Figure~\ref{cmb_ani}) the  relative difference increases as $p$ increases. However, when considering $N_{\star}=60$ (bottom) the relative difference is lower and the best fit to CMB data, is given by the case $p=4$.

As we mention above, larger values of $N_{\star}$ are most likely according to CMB data. Thus, the correlation between $N_{\star}$ and $n_s$ implies that for the spectral index the most likely values are larger as well. However, notice that the range of maximum probability of $n_s$ obtained for each $p$ (see dark regions ($1\sigma$) in the $N_{\star}-n_s$ plane in Figure~\ref{post}) is all contained within the standard deviation reported by Planck Satellite Collaboration, given by $n_s = 0.965\pm 0.004$.

\section{Discussion}\label{con}

The inflationary paradigm stipulates an exponential expansion at the initial stage of the universe, and the signatures of such process can be inferred from the CMB radiation. Particularly, the presence of primordial gravitational waves is a prediction of inflation and, if detected, this will allow to understand the physics of the universe at very early times. In the present work we have analyzed the family of models dubbed as $\alpha$-attractors in the form proposed in ref.~\cite{German:2021rin}. While in previous works such models have been explored semi-analytically, showing a favourable behavior to describe the physics of inflation (and reheating), we now have implemented, through numerical and statistical analyses, their viability as promising candidates for an inflaton potential in the light of CMB data.

It is interesting to note that the value of the tensor-scalar ratio $r$ remains approximately equal to $0.0025$ regardless of the values of the parameter $p$ for the four cases studied. This value of $r$ falls comfortably within the upper limits imposed by current collaborations such as BICEP/Keck ($r < 0.032$) and the expected values of future experiments such as LiteBIRD ($r < 0.003$). It would be interesting to study cases with $p>4$ to see whether the value $r\approx 0.0025$ starts to change significantly (work is being done in that direction).

\begin{figure}[h!]
\centering 
\includegraphics[width=0.7\textwidth]{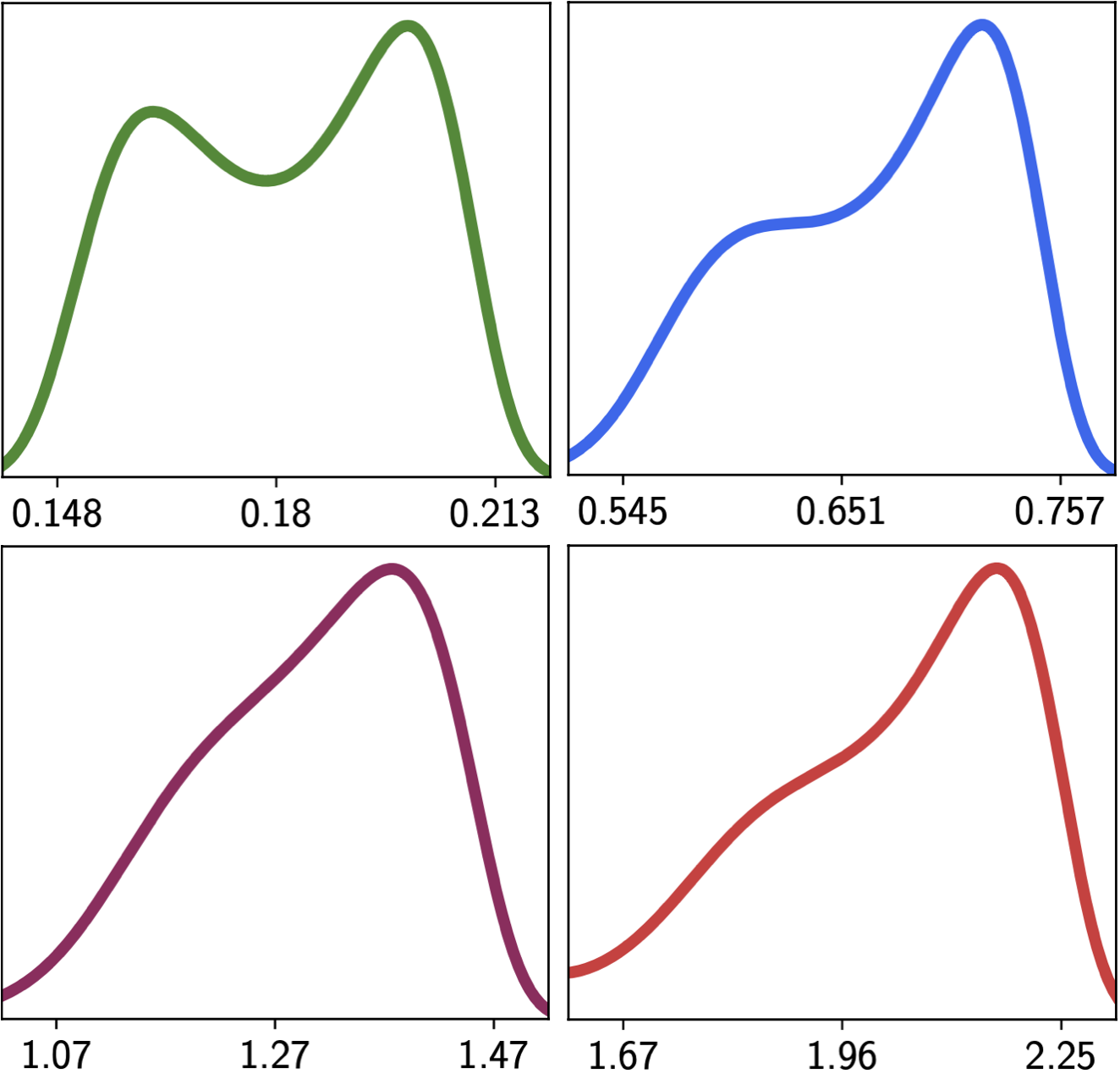}
\caption{\label{alpha_posteriors} Posteriors for the inflaton potential parameter $\alpha$ (related to the posterior of $\lambda$ through $\lambda \equiv 1/\sqrt{6\alpha}$). Top left: $p=1$ (green), top right: $p=2$ (blue), bottom left: $p=3$ (purple), and bottom right: $p=4$ (red).}
\end{figure}

Moreover, we have imposed constraints on the inflaton parameter $\lambda$, that is, after we ran chains in a MCMC process in order to find the most likely value of $\lambda$ given the CMB data. This goes beyond the superposition of semi-analytical solutions on posterior plots of the $n_s-r$ plane. The latter procedure, may give some idea on the range of values for inflationary models consistent with observations. However, the rigorous approach in order to obtain robust results is to implement a statistical analysis where the parameters of the inflationary potential of interest are properly contained in the parameter space, as done here.
As described in Section~\ref{sec:intro}, $\lambda \equiv 1/\sqrt{6\alpha}$ is related to the curvature scalar in the field space, and the values we have obtained from the posteriors for $\lambda$ (see Figure~\ref{post} and Table~\ref{lambda_values}), are useful to generate posteriors for $\alpha$, which we show in Figure~\ref{alpha_posteriors}. A reference work performing MCMC analyses for $\alpha$-attractor models, although for other inflaton potential (specifically, the double-well inflationary potential) is~\cite{Rodrigues:2020fle}.

It will be interesting to study higher values of $p$ aiming for insights related with the Bayes's factor for this family of models. From Table~\ref{Jeffrey_alpha} we see that odd values of $p$ are less favoured for increasing $p$, whereas for even $p$ the Bayes's factor is larger as long as $p$ increases, resulting in a preferred model for the larger studied value of $p$. Exploring larger values of $p$ we hope to find a maximum in the Bayes's factor. Work in progress gives us hope since we find that consistent semi-analytical solutions are obtained only if $r$ is larger than its upper bound, which should be disfavoured by the Bayesian analysis. Moreover, fractional values of $p$ can also contribute to understand whether some trend is present in the Bayes's factor for these inflationary models. This will be studied in detail elsewhere.

In summary, we have performed a Bayesian study of models belonging to a class of $\alpha$-attractors defined by the potential $V(\phi)=V_0\left[1-\text{sech}^{p}\left(\phi/\sqrt{6\alpha}M_{pl}\right)\right]$ where $\phi$ is the inflaton field and the parameter $\alpha$ corresponds to the inverse curvature of the scalar manifold in the conformal or superconformal realizations of the attractor models. We have studied in particular the cases $p=1,\, 2,\, 3\,\mathrm{and}\, 4$, with the interesting result that the tensor-to-scalar ratio $r$ takes approximately the same value $r \simeq 0.0025$ in all case. A possibility is that this value of $r$ is being reached in some ``universal'' limit (similar to what happens with the $\tanh^p$ attractors) where the power $p$ does not play an important role, and thus such parameter is useless to choose a realization preferred by data. To make this choice, we determine the Bayes factor for each case, using $p=2$ as the base model, and found that the $p=4$ case is preferred by the CMB observations, as shown in the Table~\ref{Jeffrey_alpha}. Note that the power $p$ in the potential, when expanded around the origin according to the Eq.~(\ref{potexpanded}), appears only as a multiplicative factor of the quadratic and higher order terms. Thus, the selection made by the statistical analysis is not among potentials with a radically different behaviour around the minimum (as in the $\tanh^p$ potential of Eq.~(\ref{potanh})). Instead, it is among quadratic potentials around their minimum with slightly different coefficients.


\acknowledgments

F.X.L.C.~acknowledges Proyecto CONACYT Ciencia de Frontera CF 2019/2558591 for financial support, and the computing facilities at the Laboratorio de Inteligencia Artificial y Superc\'omputo, IFM-UMSNH. G.G.~and J.C.H.~acknowledge support from program UNAM-PAPIIT, grants
IN107521 ``Sector Oscuro y Agujeros Negros Primordiales'' and IG 102123 "Laboratorio de Modelos y Datos (LAMOD) para proyectos de Investigaci\'on Cient\'ifica: Censos Astrof\'isicos".



\bibliographystyle{plunsrt}
\bibliography{bib}







\end{document}